\documentclass[12pt]{article}

\global\arraycolsep=1pt

\usepackage{color,amsbsy,amssymb,latexsym,amsfonts, amsmath,cancel}
\usepackage{braket}
\usepackage{mathrsfs}
\usepackage{graphicx}
\usepackage{cite}

\numberwithin{equation}{section}
\newcommand{\bel}[1]{\begin{equation}\label{#1}}                     
\newcommand{\bal}[1]{\begin{eqnarray}\label{#1}}   
\newcommand{\be}{\begin{equation}}               
\newcommand{\ba}{\begin{eqnarray}}           
\newcommand{\ee}{\end{equation}}
\newcommand{\ea}{\end{eqnarray}}

\renewcommand{\thefootnote}{\fnsymbol{footnote}}

\newcommand{\bea}{\begin{equation}}
\newcommand{\eea}{\end{equation}}
\newcommand\fD{\mathfrak D}

\begin{document}

\date{today}
%
%
\begin{titlepage}
\begin{flushright}
\normalsize
~~~~
OCU-PHYS 413\\
November 5, 2014
\end{flushright}

\vspace{15pt}

\begin{center}
{\Large\bf 
  Mass Sum Rule of the Effective Action  \\
   \vspace*{2mm}
  on Vacua with Broken Rigid ${\cal N} =1$  Supersymmetry }
\end{center}

\vspace{23pt}

\begin{center}
{\large H. Itoyama}$^{a, b} $
  and 
{\large Nobuhito Maru$^a$
}\\
%
\vspace{18pt}
%

$^a$ \it Department of Mathematics and Physics, Graduate School of Science\\
Osaka City University  and\\
\vspace{5pt}

$^b$ \it Osaka City University Advanced Mathematical Institute (OCAMI) \\

\vspace{5pt}

3-3-138, Sugimoto, Sumiyoshi-ku, Osaka, 558-8585, Japan \\

\end{center}
%
\vspace{20pt}
\begin{center}
{\bf Abstract} \\
\end{center}
%

We present an extension of the mass sum rule that applies to renormalizable rigid supersymmetric field theories 
 to the case of the ${\cal N} =1$ supersymmetric effective action (the gauged non-linear sigma model) 
 consisting of adjoint scalar superfields and vector superfields 
 possessing a K\"ahler potential, a set of gauge coupling functions (second prepotential derivatives) 
 and a superpotential, which respectively set their energy scales. 
The mass sum rule derived is valid for any vacua, 
 including the (metastable) one of broken supersymmetry with the condensates of $D$-term and/or $F$-term.
We manage to extend these analyses to the cases where superfields in (anti-)fundamental representation are present.
The supertrace is shown to vanish in those cases 
 where underlying geometry is special K\"{a}hler and theory under concern is anomaly free.
Simple phenomenological application is given, providing an upper bound for gaugino masses.
 We discuss that the effects of the $D$ and/or $F$ condensates can be represented 
 as a set of soft breaking terms with their strengths predicted by the scales.


\vfill

\setcounter{footnote}{0}
\renewcommand{\thefootnote}{\arabic{footnote}}

\end{titlepage}

\date{today}
\renewcommand{\thefootnote}{\arabic{footnote}}
\setcounter{footnote}{0}

\section{Introduction}

  The mass sum rule of renormalizable rigid supersymmetric (SUSY) field theories in four dimensions \cite{sumrule}
  played an important role
 in eighties in  deciding upon the appropriate use of supersymmetry in particle physics together with
 the notion of  naturalness. Being largely independent of the dynamics, it gives us a general constraint on
   a pattern of bose-fermi mass splitting when applied to theories with vacua of spontaneously  broken supersymmetry
  and has provided a rationale for the existence of the hidden sector that
 has affected the SUSY model building till today.
 After the three decades, nature appears to call for a renewed version of naturalness 
  while supersymmetry has been confronted with more and more stringent bounds from the experiments \cite{SUSYATLAS, SUSYCMS}. 
   
 The effective action is an appropriate tool to summarize quantum properties of field theoretic system
  seen as low energy dynamics: its form is dictated by the symmetries of the system and
  the coefficient functions represent quantum effects of ``the high frequency part" integrated over 
  (see, for instance, \cite{WK,Peskin}).

 In this paper, we derive a mass sum rule from a prototypical ${\cal N} =1$ supersymmetric effective action
    (gauged non-linear sigma model).
  The effective action that we consider consists of
    adjoint chiral superfields and vector superfields, possessing a K\"{a}hler potential, a set of gauge coupling functions
(second prepotential derivatives) and a superpotential, which respectively set distinct energy scales. 
 Deriving the mass sum rule of this system is interesting as the system incorporates naturally the notion of Dirac gaugino 
  or Majorana-Dirac gaugino scenario which is receiving attention recently as an extension of the spectrum 
  in the MSSM gauge sector \cite{Fayet, PS, HR, FNW, NRSU, CFM, CFK, ABDQT, NPT, ABDQ, HNNS, Hsieh, 
  KPW, BN, KMR, ABFP, BG, Blechman, Carpenter, KOR, AG, DMRMC, BGM, FG, KM, Goodsell, Unwin, 
  BGS, MNS, AB, DGHT, IMaru1, IMaru2, IMaru3, IMaru4}. 
The sum rule, as is always the case, represents the symmetry of the action, being insensitive to the structure or the choice of vacua. 
The real interest in the supersymmetric sum rule
  lies, of course, in those cases where the bose-fermi degeneracy of the spectrum is lifted.
 It has been demonstrated that dynamical supersymmetry breaking takes place on metastable vacua
    in the weak-coupling regime: the D-term triggered Hartree-Fock treatment has enabled us to exhibit the condensates
     of the order parameters of supersymmetry on the metastable vacuum through the gap equation \cite{IMaru1, IMaru2, IMaru3}.
  The fields in the observable sector pick the effects of these condensates through
   the tree level analysis of the effective action.  The application of the sum  rule we derive is, however, not 
   going to be limited to this particular situation.

 In the next section, we recall the effective action mentioned above.
The scales are contained in the three input functions. We consider the (metastable) vacua which break
 supersymmetry.
In section 3, we introduce the boson and fermion mass matrices and compute matrix elements.  
In section 4, we derive the mass sum rule from the matrix elements, temporarily assuming unbroken
 gauge symmetry.  
It is shown that the supertrace of the mass matrices squared vanishes 
 in those cases where the underlying geometry is  special K\"ahler.  
In section 5, we extend our analyses to the  general case where the gauge symmetry is broken 
 and (anti-)fundamental matter superfields
 are included.  We complete the derivation of the mass sum rule to this general case.
 The right hand side of the mass sum rule vanishes by the special K\"ahler geometry one adopts
  and the anomaly free property of the theory under concern.
 In section 6, we present a simple application of the mass sum rule, which leads to an allowed
  range of the gluino mass. 
 The terms generated by $D$ and $F$ condensates (or stationary values) can, in practice, 
 be recognized as a set of soft breaking terms, using the spurion technique \cite{GG}. 
 We exhibit these in section 7.
 Throughout the paper, we work with the notation, so that our computation and results are
 insensitive to the vacua one explores.
 In the appendix, we touch upon how expressions such as the matrix elements get further converted
 in some simplest cases.

\section{ ${\cal N}=1$ effective action of adjoint scalar and vector superfields}

Let us first consider the general ${\cal N}=1$ supersymmetric action consisting of chiral superfield $\Phi^a$ 
   in the adjoint representation and the vector superfield $V^a$:
  \ba
    {\cal L}
     &=&     
           \int d^4 \theta K(\Phi^a, \bar{\Phi}^a) + (gauging)   
        + \int d^2 \theta
           {\rm Im} \frac{1}{2} 
           \tau_{ab}(\Phi^a)
           {\cal W}^{\alpha a} {\cal W}^b_{\alpha}
            + \left(\int d^2 \theta W(\Phi^a)
         + c.c. \right).   \nonumber \\    
           \label{theaction}
    \ea  
There are three input functions: 
 the K\"{a}hler potential $K(\Phi^a, \bar{\Phi}^a)$ with its gauging, 
 the gauge kinetic superfields $\tau_{ab}(\Phi^a)$ 
 that are the second derivatives of a holomorphic function ${\cal F}(\Phi^a)$,
 and a superpotential $W(\Phi^a)$.

 In parallel to \cite{IMaru3, IMaru4}, we postulate the followings: 
\begin{itemize}
 \item[1)] third derivatives of ${\cal F}(\Phi^a)$ at the scalar VEV's are non-vanishing. 
 \item[2)] the superpotential at tree level preserves ${\cal N}=1$ supersymmetry.  
 \item[3)] the gauge group can be arbitrary except that it contains an overall $U(1)$ 
               in which all particles in the observable sector are singlets.
 \end{itemize}
 It has been demonstrated \cite{IMaru3}
 the supersymmetry is spontaneously broken in the Hartree-Fock approximation in this system,
  replacing $3)$ by
 \begin{itemize}
\item[3)'] the vacuum is taken to be in the unbroken phase of the gauge group, which is taken to be
  $U(N)$ for definiteness. 
\end{itemize}
This last assumption has been made for a technical reason. 
   
There are, in principle, three scales in accordance with the three input functions.
In order to avoid complications which are unnecessary in this paper, 
 we consider the case in which the scale set by the K\"ahler potential 
 and the one set by the prepotential are the same order. 
 (This is automatically satisfied in the special K\"ahler case or practically correct 
 in the case where the $D^0$ condensate is dominant against  the $F^0$ condensate.)  
The one of the two fundamental scales is, therefore, taken to be
 the mass parameter $M_{prep}$ contained in the prepotential function ${\cal F}$. 
The other is the mass parameter $M_{sup}$ contained in the superpotential $W$. 
The SUSY breaking scale, namely, the order parameter $\langle D^0 \rangle$ 
 is found to be given by their geometric mean 
$\langle D^0 \rangle \sim M_{prep} M_{sup}$. (See eq.(3.13) of \cite{IMaru3} for the derivation). 
The susy breaking scale can, therefore, be arbitrarily large, depending upon how large these two parameters are. 
All of the adjoint multiplets of the standard model group appearing in our theory receive mass of order $M_{sup}$.        
The role played by this effective action in the vacuum of broken supersymmetry is somewhat analogous
to that played by the NJL model \cite{NJL,BCS} in broken chiral symmetry which connects the confinement scale 
and the scale of the chiral lagrangian: here, this effective action  describes
 the dynamics  in the intermediate energy scale, connecting the
 low energy dynamics with the high energy inputs.

\section{Mass matrices and computation of the matrix elements}
In this section and the next section, we present the principal part of our computation. For the sake of our presentation,
  we temporarily limit ourselves to the case of unbroken gauge group, ignoring  spin-one contribution  as well
   as additional scalar-scalar and $D$-scalar contributions to mass matrices due to eq.(\ref{killingptl}). 
   These can be readily put in, which we will do in section 5 where we consider the general case that includes 
   the broken gauge group and matter supermultiplets.

Let us study  the quadratic fluctuations of the action around its stationary points 
 of the scalar fields and the auxiliary fields. This leads us to mass formulas of the effective action 
 on a generic vacuum of dynamically broken ${\cal N}=1$ supersymmetry. 
We begin by separating the stationary values (VEV's) of the scalar fields, 
 the auxiliary fields, denoted by $\varphi_*^a$ and by $D_*^a$ and $F_*^a$ respectively, 
 from their fluctuations: 
\begin{eqnarray}
&&{\cal L}(\Phi_*^a+\Phi^a, V_*^a+V^a) = {\cal L}(\Phi_*^a, V_*^a) 
+ {\cal L}_{fluc}(\Phi^a, V^a;\Phi_*^a, V_*^a), \\
&&\Phi_*^a = \varphi_*^a + \theta \theta F_*^a, \quad V_*^a  = \frac{1}{2} \theta \theta \bar{\theta} \bar{\theta} D_*^a. 
\end{eqnarray}

The terms in ${\cal L}_{fluc}$ which are quadratic in fluctuations can be represented as
\ba
&&{\cal L}^{quad} = {\cal L}_B^{quad} + {\cal L}_F^{quad}, \quad 
{\cal L}_B^{quad} \equiv K_B - V_{2B}, \quad {\cal L}_F^{quad} \equiv K_F - V_{2F}, \\
&& K_B = g_{a\bar{b}*} \partial_\mu \phi^a \partial^\mu \bar{\phi}^{\bar{b}}
  - \frac{1}{4} (\Im {\cal F})_{ab} F_{\mu \nu}^a F^{b \mu \nu} , \quad 
g_{a\bar{b}*} \equiv g_{a\bar{b}} (\varphi_*^c, \bar{\varphi}_*^{\bar{c}}) = g_{\bar{b}a*}, \\
&&K_F = -\frac{i}{2}g_{ab*} \psi^a\sigma^\mu \partial_\mu \bar{\psi}^b + \frac{i}{2} g_{ab*}(\partial_\mu \psi)^a \sigma^\mu \psi^b 
-\frac{1}{2} {\cal F}_{ab*} \lambda^a \sigma^\mu \partial_\mu \bar{\lambda}^b 
- \frac{1}{2} \bar{\cal F}_{ab*} \partial_\mu \lambda^a \sigma^\mu \bar{\lambda}^b, \\
&&{\cal F}_{ab*} \equiv {\cal F}_{ab}(\varphi_*^c), \\
&&V_{2B} = \frac{1}{2} (\bar{\phi}, \phi, D, \bar{F}, F) {\cal M}_B^2 
\left(
\begin{array}{c}
\phi \\
\bar{\phi} \\
D \\
F \\
\bar{F}
\end{array}
\right), \label{V2B} \\
&&V_{2F} = 
\frac{1}{2} (\lambda, \psi) {\cal M}_F
\left(
\begin{array}{c}
\lambda \\
\psi \\
\end{array}
\right) 
+ 
\frac{1}{2} (\bar{\lambda}, \bar{\psi}) \overline{{\cal M}}_F
\left(
\begin{array}{c}
\bar{\lambda} \\
\bar{\psi} \\
\end{array}
\right). 
\label{V2F}
\ea
Here in eq.(\ref{V2B}) and eq.(\ref{V2F}), we have adopted matrix notation which 
is self-explanatory: the adjoint indices $a,b, \cdots$ have been suppressed.
The matrices ${\cal M}_B^2$, ${\cal M}_F$ and $\overline{{\cal M}}_F$ consist of blocks of matrices of 
smaller size and are displayed as
\ba
&&{\cal M}_B^2 = 
\left(
\begin{array}{ccccc}
{\cal M}_{\bar{\phi} \phi}^2 & {\cal M}_{\bar{\phi} \bar{\phi}}^2 & {\cal M}_{\bar{\phi}D}^2& {\cal M}_{\bar{\phi}F}^2 & {\cal M}_{\bar{\phi}\bar{F}}^2 \\
{\cal M}_{\phi \phi}^2 & {\cal M}_{\phi \bar{\phi}}^2 & {\cal M}_{\phi D}^2 & {\cal M}_{\phi F}^2 & {\cal M}_{\phi \bar{F}}^2 \\
{\cal M}_{D \phi}^2 & {\cal M}_{D \bar{\phi}}^2 & {\cal M}_{DD}^2 & 0 & 0 \\
{\cal M}_{\bar{F} \phi}^2 & {\cal M}_{\bar{F} \bar{\phi}}^2 & 0 & {\cal M}_{\bar{F}F}^2 & 0 \\
{\cal M}_{F \phi}^2 & {\cal M}_{F \bar{\phi}}^2 & 0 & 0 & {\cal M}_{F\bar{F}}^2 \\
\end{array}
\right), \\
&&{\cal M}_F = 
\left(
\begin{array}{cc}
{\cal M}_{\lambda \lambda} & {\cal M}_{\lambda \psi} \\
{\cal M}_{\psi \lambda} & {\cal M}_{\psi \psi} \\
\end{array}
\right), 
\quad 
\overline{{\cal M}}_F = 
\left(
\begin{array}{cc}
\bar{{\cal M}}_{\bar{\lambda} \bar{\lambda}} & \bar{{\cal M}}_{\bar{\lambda} \bar{\psi}} \\
{\cal M}_{\bar{\psi} \bar{\lambda}} & {\cal M}_{\bar{\psi} \bar{\psi}} \\
\end{array}
\right). 
\ea

We have computed the entries of these three matrices and they  are respectively given as
\ba
 - {\cal M}_{B}^2
&=& 
\left(
\begin{array}{c|c|c|c|c}
(F \cdot \bar{\partial} \partial g \cdot \bar{F})_* & (F \cdot \bar{\partial} \bar{\partial} g \cdot \bar{F})_* & -\frac{((\bar{\partial} \bar{{\cal F}} \cdot ) D)_*}{2i} 
 & ((\bar{\partial} g \cdot ) \bar{F})_* & ((\bar{\partial} g \cdot ) F)_* \\
 & + ((\bar{F} \cdot ) \bar{\partial} \bar{\partial} \overline{(\partial W)})_* & & & +(\bar{\partial} \bar{\partial} \bar{W})_* \\
 & -\frac{(D \cdot \bar{\partial} \bar{\partial} \bar{\cal F} \cdot D)_*}{4i} & & & \\
\hline
(F \cdot \partial \partial g \cdot \bar{F})_* & (F \cdot \partial \bar{\partial} g \cdot \bar{F})_* 
 & \frac{((\partial {\cal F} \cdot )D)_*}{2i} & ( (\partial g \cdot ) \bar{F})_* & ( (\partial g \cdot ) F)_* \\
+ ( ( F \cdot ) \partial \partial(\partial W))_*  & & & +(\partial \partial W)_* & \\
+ \frac{(D \cdot \partial \partial {\cal F} \cdot D)_*}{4i} &  &  &  & \\
\hline
\frac{((\partial {\cal F} \cdot ) D)_*}{2i} & - \frac{((\bar{\partial} \bar{{\cal F}} \cdot )D)_*}{2i} & ({{\rm Im}{\cal F}})_* & 0 & 0 \\
\hline
((\partial g \cdot ) F)^T_* & ((\bar{\partial} g \cdot ) F)^T_* & 0 & g_* & 0 \\
 & +(\bar{\partial} \bar{\partial} \bar{W})_* & & & \\
\hline
((\partial g \cdot ) \bar{F})^T_* & ((\bar{\partial} g \cdot ) \bar{F})^T_* & 0 & 0 & g_* \\
+(\partial \partial W)_* & & & & \\
\end{array}
\right), 
\label{bosonmass} \\
{\cal M}_F 
&=&
\left(
\begin{array}{c|c}
-\frac{i}{2} (\partial {\cal F} \cdot) F & -\frac{\sqrt{2}}{4} (\partial {\cal F} \cdot) D \\
\hline
-\frac{\sqrt{2}}{4} (\partial {\cal F} \cdot) D & \partial \partial W + (\partial g \cdot ) \bar{F} \\
\end{array}
\right)_*, 
\label{fermionmass} \\
\overline{{\cal M}}_F 
&=&
\left(
\begin{array}{c|c}
\frac{i}{2}  (\bar{\partial} \bar{{\cal F}} \cdot) \bar{F} & -\frac{\sqrt{2}}{4} (\bar{\partial} \bar{{\cal F}} \cdot) D \\
\hline
-\frac{\sqrt{2}}{4} (\bar{\partial} \bar{{\cal F}} \cdot) D & \bar{\partial} \bar{\partial} \bar{W} + (\bar{\partial} g \cdot) F \\
\end{array}
\right)_*. 
\label{fermionmassbar}
\ea
Here again, we have introduced a shorthand notation: for instance,  
$\left. (F \cdot \bar{\partial} \partial g \cdot \bar{F})_* \right._{\bar{a} b} =
F^c_*  \bar{\partial}_{\bar{a} } \partial_{b} g_{c \bar{c} *}  \bar{F}^{\bar{c}}_*$
 as well as 
 $\left.((\partial {\cal F} \cdot ) D)_* \right._{ab} = {\cal F}_{abc*} D^c_*$.
The notation is generic, so that our computation in what follows and the mass sum rule in the next section
 are insensitive to the structure/pattern of vacua explored. 
For an example of the expressions at a specific vacuum, see the Appendix.


 Note that we did not include here the contributions from the killing potential
\ba
\fD_a = -\frac{1}{2}({\cal F}_b f_{ac}^b \bar{\phi}^c + \bar{{\cal F}}_b f_{ac}^b \phi^c). 
\label{killingptl}
\ea
(See, for instance, \cite{FIS1,IMM}.)
For the boson mass term, the term in the action attendant with eq.(\ref{killingptl}) is a generalization of 
the scalar potential due to gauge interactions
in the renormalizable SUSY gauge theories:
\ba
\frac{1}{2} (D_*^a + D^a) \fD_a (\varphi_* + \phi, \bar{\varphi}_* + \bar{\phi}). 
\label{killingbosonmass}
\ea
For the fermion mass term, the term attendant with eq.(\ref{killingptl})  is
\ba
&&\frac{1}{\sqrt{2}} g_{ab*} (\lambda^c \psi^a \bar{k}_{c*}^b + \bar{\lambda}^c \bar{\psi}^b k_{c*}^a), 
\label{killingfermionmixing} \\
&& k_a^b = -i g^{bc} \bar{\partial}_c \fD_a. 
\label{killingvector}
\ea
The killing potential
$\fD_a$ contains the structure constant as a multiplicative factor and 
these terms do not contribute to the mass matrices  in the unbroken phase of the gauge group. 
We will put these back in section 5. 

The quadratic form eq.(\ref{V2B}) can be simplified by ``completing the square" for the auxiliary fields: 
\ba
V_{2B} = \frac{1}{2} (\bar{\phi}, \phi, D', \bar{F}', F') {\cal M}_{B,red}^2 
\left(
\begin{array}{c}
\phi \\
\bar{\phi} \\
D' \\
F' \\
\bar{F}'
\end{array}
\right),
\ea
\ba
{\cal M}_{B, red}^2 = 
\left(
\begin{array}{ccccc}
{\cal M}_{red~\bar{\phi} \phi}^2 & {\cal M}_{red~\bar{\phi} \bar{\phi}}^2 & 0 & 0 & 0 \\
{\cal M}_{red~\phi \phi}^2 & {\cal M}_{red~\phi \bar{\phi}}^2 & 0 & 0 & 0 \\
0 & 0 & {\cal M}_{DD}^2 & 0 & 0 \\
0 & 0 & 0 & {\cal M}_{\bar{F}F}^2 & 0 \\
0 & 0 & 0 & 0 & {\cal M}_{F\bar{F}}^2 \\
\end{array}
\right). 
\label{redM2}
\ea
Here 
\ba
{\cal M}_{red~AB}^2 &=& {\cal M}_{AB}^2
 - \sum_{\alpha, \beta=D, F, \bar{F}} {\cal M}_{A\alpha}^2 ({\cal M}^2)^{-1}_{\alpha \beta} {\cal M}_{\beta B}^2, 
\label{red} \\
D' &=& D + ({\cal M}_{DD}^2)^{-1} {\cal M}_{D \phi}^2 \phi +  ({\cal M}_{DD}^2)^{-1} {\cal M}_{D \bar{\phi}}^2 \bar{\phi}, \\
F' &=& F + ({\cal M}_{F\bar{F}}^2)^{-1} {\cal M}_{\bar{F} \phi}^2 \phi +  ({\cal M}_{F\bar{F}}^2)^{-1} {\cal M}_{\bar{F} \bar{\phi}}^2 \bar{\phi}, \\
\bar{F}' &=& \bar{F} + ({\cal M}_{\bar{F}F}^2)^{-1} {\cal M}_{F \phi}^2 \phi +  ({\cal M}_{\bar{F}F}^2)^{-1} {\cal M}_{F \bar{\phi}}^2 \bar{\phi}. 
\ea

\section{Mass sum rule}

Our consideration in the last section is enough to lead us to the mass sum rule 
 for the class of supersymmetric effective field theories that we consider in this paper. 
It is a generalization of the well-known sum rule \cite{sumrule}
 which applies for the models of supersymmetric field theories with canonical kinetic terms 
 in the sense that eq.(\ref{theaction}) contains the K\"ahler potential 
 and the gauge coupling function (the prepotential derivatives) as well. 

In the vacua where the gauge group is unbroken, the gauge bosons are massless 
 and the scalar masses are obtained by diagonalizing 
 \ba
 \sqrt{g_*^{-1}}
 \left(
 \begin{array}{cc}
 {\cal M}_{red~\bar{\phi} \phi}^2 & {\cal M}_{red~\bar{\phi} \bar{\phi}}^2 \\
 {\cal M}_{red~\phi \phi}^2 & {\cal M}_{red~\phi \bar{\phi}}^2 \\
   \end{array}
 \right)
 \sqrt{g_*^{-1}}. 
 \ea
The sum of the boson masses squared is, therefore, given by
\ba
{\rm Tr} M^2_{bosons} \equiv tr g_*^{-1} {\cal M}_{red~\bar{\phi}\phi}^2 +  tr g_*^{-1} {\cal M}_{red~\phi \bar{\phi}}^2. 
\ea
Using eq.(\ref{bosonmass}) and eq.(\ref{red}) , we obtain
\ba
{\rm Tr}M^2_{bosons} &=& 
tr \left[
-2(g^{-1}F \cdot \bar{\partial} \partial g \cdot \bar{F})_* 
 + \frac{1}{2} g_*^{-1} ((\bar{\partial} \bar{{\cal F}} \cdot)D)_* ({\rm Im}{\cal F})^{-1}_* ((\partial {\cal F} \cdot) D)_* 
 \right. \nonumber \\
 &&\left. 
 + 2(g^{-1} \bar{\partial} g \cdot \bar{F})_* g_*^{-1} ( \partial g \cdot F)_*^T
 + 2 ((g^{-1}  \partial g \cdot \bar{F})_* + (g^{-1} \partial \partial W)_*)
 (g^{-1}_* (\bar{\partial} g \cdot F)^T_* 
 \right. \nonumber \\
 &&\left.   
 + g^{-1} (\bar{\partial} \bar{\partial} \bar{W})_*)
\right]
\ea
Here $tr$ denotes the sum over the adjoint indices. 

As for fermion masses, they are obtained by diagonalizing
\ba
\sqrt{G_{F*}^{-1/2}} {\cal M}_F \sqrt{G_{F*}^{-1/2}}~{\rm or}~\sqrt{G_{F*}^{-1/2}} \overline{{\cal M}_F}\sqrt{G_{F*}^{-1/2}},  
\ea
where
\ba
G_{F*} = 
\left(
\begin{array}{c|c}
({\rm Im}{\cal F})_* & 0 \\
\hline
0 & g_* \\
\end{array}
\right). 
\ea
The sum of the fermion masses squared including the factor 2 
 due to the number of polarizations per particle is given by
\ba
2 {\rm Tr} M^2_{fermions} &\equiv& 
 tr \left[ {\cal M}_F G_{F*}^{-1} \overline{{\cal M}_F} G_{F*}^{-1} \right] +
   tr \left[  \overline{{\cal M}_F}  G_{F*}^{-1}   {\cal M}_F  G_{F*}^{-1}  \right]   \\
&=& tr 
\left[
\frac{1}{2} ((\partial {\cal F} \cdot)F)_* ({\rm Im}{\cal F})^{-1}_* ((\bar{\partial} \bar{{\cal F}} \cdot)\bar{F})_* ({\rm Im}{\cal F})^{-1}_*
+ \frac{1}{2} ({\rm Im}{\cal F})^{-1}_* ((\partial{\cal F} \cdot) D)_* g_*^{-1} ((\bar{\partial} \bar{{\cal F}} \cdot) D)_* 
\right. \nonumber \\
& & \left. 
+ 2 (\partial \partial W + \partial g \cdot \bar{F})_* g_*^{-1}  (\bar{\partial} \bar{\partial} \bar{W} + \bar{\partial} g \cdot F)_* g_*^{-1}
\right]. 
\ea
Hence we obtain
\ba
{\rm Tr} M^2_{bosons} -2 {\rm Tr} M^2_{fermions}
&=& 
tr \left[
-2(g^{-1}F \cdot \bar{\partial} \partial g \cdot \bar{F})_* 
 + 2(g^{-1} (\bar{\partial} g \cdot )\bar{F})_* g_*^{-1} ((\partial g \cdot) F)_*^T
\right. \nonumber \\
&& \left. 
- \frac{1}{2} ({\rm Im}{\cal F})^{-1}_* ((\partial {\cal F} \cdot)F)_* ({\rm Im}{\cal F})^{-1}_* ((\bar{\partial} \bar{{\cal F}} \cdot) \bar{F})_* 
\right],
\label{masssumrulesec4}
\ea
observing partial cancellations. 

 This expression vanishes
in those cases where the underlying geometry is special K\"ahler, whose condition is given by
$g={\rm Im}{\cal F},$ and  $\partial \bar{\partial} g = 0$.
 
\section{The general case of broken gauge group and inclusion of matter multiplets in the (anti-)fundamental representation}

So far, we have dealt with those cases where only the matter chiral multiplets in the adjoint representation are present. 
In order to confront our analysis with more realistic particle spectrum and patterns, we need to work with cases 
 with broken gauge symmetry and where the matter chiral multiplets in the (anti-)fundamental representation are present. 
In this section, as one of the prototypical examples, we add to the original Lagrangian 
 the one consisting of a pair of chiral superfields $(H^i, H_{c i_c})$ 
 belonging to the fundamental and the anti-fundamental representations respectively: 
\ba
{\cal L}_f = \int d^4 \theta 
\left(
\bar{H} e^{V} H + \bar{H}_c e^{-V} H_c
\right). 
\label{matterlagrangian}
\ea
The superfields are expanded as
\ba
&&H = h(y) + \sqrt{2} \theta \psi_h(y) + \theta \theta F_h(y), \\
&&H_c = h_c(y) + \sqrt{2} \theta \psi_{h_c}(y) + \theta \theta F_{h_c}(y)
\ea
with $y^\mu \equiv x^\mu +i\theta \sigma^\mu \bar{\theta}$
and their stationary values are denoted by
\ba
H_* = h_* + \theta \theta F_{h*}, \quad H_{c*} = h_{c*} + \theta \theta F_{h_c*}. 
\ea
The superpotential term is appropriately extended to include these matter chiral multiplets as well: 
\ba
{\cal L}_{sup}^{extended} &=& F_\phi^a \partial_a W 
+ F_h^i \partial_i W + F_{h_c i_c} \partial^{i_c} W \nonumber \\
&&-\frac{1}{2} \sum_{A=a,i,i_c} \sum_{B=b,j,j_c} 
\left(
\psi_\phi^{A=a}, \psi_h^{A=i}, \psi_{h_c A=i_c}
\right)
\partial_A \partial_B W
\left(
\begin{array}{c}
\psi_\phi^{B=b} \\
\psi_h^{B=j} \\ 
\psi_{h_c B=j_c} \\
\end{array}
\right) + {\rm c.c.} \\
&\equiv& 
F^A \partial_A W -\frac{1}{2} \psi^A (\partial_A \partial_B W) \psi^B + {\rm c.c.}. 
\ea
Here we have denoted by $A=(a, i, i_c)$ and by $B=(b, j, j_c)$ 
a collection of adjoint, fundamental, and anti-fundamental indices. 
The theory extended this way is given by the Lagrangian
\ba
{\cal L} = 
\int d^4 \theta K(\Phi^a, \bar{\Phi}^a) + (gauging) 
+ \int d^2 \theta 
 {\rm Im} \frac{1}{2} 
  \tau_{ab}(\Phi^a)
  {\cal W}^{\alpha a} {\cal W}^b_{\alpha}
           + {\cal L}_{sup}^{extended} + {\cal L}_f. 
\label{Lextended}
\ea 

Let us now turn to the computation of the matrix elements of the boson mass matrix 
 and that of the fermion mass matrix in the extended theory. 
Some of the changes we have to make as compared with the computation done in section 3 
 are just the extension of the adjoint index $a, b \cdots$ to $A=(a, i, i_c), B=(b, j, j_c) \cdots$ 
 as we have simply added species of chiral matter multiplets. 
The forms of the matrices ${\cal M}_B^2, {\cal M}_F, \overline{{\cal M}}_F$ 
 in eq.(\ref{bosonmass}), eq.(\ref{fermionmass}) and eq.(\ref{fermionmassbar}) are still relevant in this section as well 
 and we use the same symbols with the index extension understood. We just need to add
 a collection of rows and a collection of columns to ${\cal M}_B^2$ to include the spin one contribution.

There are, however, new contributions due to the fact that we work here in the vacua of broken gauge symmetry. 
By Higgs mechanism, there are massive spin one particles which gain their masses  by (a generalization) 
 of seagull interactions in the first and the last terms of eq.(\ref{Lextended}) and, therefore, 
 by  the derivatives of an appropriate generalization of the killing potential $\fD_a$. 
 The nonvanishing block is denoted by $(\Delta {\cal M}_B^2)_{V V}$. 
There are also new contributions to the matrix elements of the four blocks of the scalar-scalar part 
 and to those of another four blocks of the $D$-scalar part as well by eq. (\ref{killingbosonmass}) 
 and by $\frac{1}{2}D^a \bar{h} T^a h -\frac{1}{2} D^a \bar{h}_c T^a h_c$, which is
  obtained from eq. (\ref{matterlagrangian}). 
As for the fermion mass matrix, the new contributions are read off from eqs. (\ref{killingfermionmixing}), (\ref{killingvector}). 

Putting all these together, we write the increment of the boson mass matrix denoted by $\Delta {\cal M}_B^2$ 
 as
\ba
\Delta {\cal M}_B^2 = 
\left(
\begin{array}{cccccc}
(\Delta {\cal M}_B^2)_{\bar{\phi}\phi} & (\Delta {\cal M}_B^2)_{\bar{\phi}\bar{\phi}} & 0 & (\Delta {\cal M}_B^2)_{\bar{\phi}D} & 0 &~ 0 \\
(\Delta {\cal M}_B^2)_{\phi\phi} & (\Delta {\cal M}_B^2)_{\phi\bar{\phi}}& 0 & (\Delta {\cal M}_B^2)_{\phi D} & 0 &~ 0 \\
0 & 0 & (\Delta {\cal M}_B^2)_{V V} & 0 & 0 &~ 0 \\
(\Delta {\cal M}_B^2)_{D \phi} & (\Delta {\cal M}_B^2)_{D \bar{\phi}} & 0 & 0 & 0 &~ 0 \\
0 & 0 & 0 & 0 & 0 &~ 0 \\
0 & 0 & 0 & 0 & 0 &~ 0\\
\end{array}
\right).
\ea
Here, we have added the third row and third column as the part in
 which spin one massive particles are involved. 
The entries are computed to be 
\ba
&&(\Delta {\cal M}_B^2)_{\bar{\phi}\phi} = (\Delta {\cal M}_B^2)_{\phi \bar{\phi}} = -\frac{1}{2} D_* \cdot (\partial \bar{\partial} \hat{\fD})_*, \\
&&(\Delta {\cal M}_B^2)_{\phi \phi} =  -\frac{1}{2} D_* \cdot (\partial \partial \hat{\fD})_*, \quad 
(\Delta {\cal M}_B^2)_{\bar{\phi}\bar{\phi}}  =  -\frac{1}{2} D_* \cdot (\bar{\partial} \bar{\partial} \hat{\fD})_*, \\
&&(\Delta {\cal M}_B^2)_{\phi D} = -\frac{1}{2} (\partial \hat{\fD})_*, \quad
   (\Delta {\cal M}_B^2)_{D\phi} = -\frac{1}{2} (\partial \hat{\fD})_*^T,    \\
&&(\Delta {\cal M}_B^2)_{\bar{\phi}D}  = -\frac{1}{2} (\bar{\partial} \hat{\fD})_*, \quad 
    (\Delta {\cal M}_B^2)_{D \bar{\phi}} = -\frac{1}{2} (\bar{\partial} \hat{\fD})_*^T,   \\
&&(\Delta {\cal M}_B^2)_{VV} 
= \frac{1}{4} 
\left[
(\partial \hat{\fD})_*^T g_*^{-1} (\bar{\partial} \hat{\fD})_* + (\bar{\partial} \hat{\fD})_*^T g_*^{-1} (\partial \hat{\fD})_* 
\right]. 
\ea
Here we have denoted by $\hat{\fD}$ the killing potential appropriately extended to include the contributions 
 from, $h, \bar{h}, h_c$ and $\bar{h}_c$,
\ba
\hat{\fD}^a = \fD^a +  (\bar{h} T^a h - \bar{h}_c T_c^a h_c). 
\ea 
We have also made an index extension of the K\"{a}hler metric
\ba
g_{a\bar{b}}   \Rightarrow g_{A\bar{B}} = 
\left(
\begin{array}{ccc}
g_{a \bar{b}} & 0 & 0 \\
0 & \delta_i^{~\bar{i}} & 0 \\
0 & 0 & \delta_{~\bar{i_c}}^{i_c} \\
\end{array}
\right).
\ea
From these data, the increment of ${\cal M}_{B, red}^2$ from eq. (\ref{redM2}) to the current case is
\ba
(\Delta{\cal M}_{B, red}^2)_{AB} &\equiv& 
\left({\cal M}_{B, red}^2({\cal M}_B^2+\Delta{\cal M}_B^2) \right)_{AB}  - 
\left({\cal M}_{B, red}^2({\cal M}_B^2) \right)_{AB} 
\nonumber \\
&=& 
(\Delta{\cal M}_{B}^2)_{AB} 
- \sum_{\alpha, \beta = D,F,\bar{F}} 
\left\{
(\Delta{\cal M}_{B}^2)_{A \alpha}
({\cal M}_{B}^2)_{\alpha \beta}^{-1} ({\cal M}_{B}^2)_{\beta B}
\right. \nonumber \\
&& \left. + ({\cal M}_{B}^2)_{A \alpha} ({\cal M}_{B}^2)_{\alpha \beta}^{-1} (\Delta{\cal M}_{B}^2)_{\beta B} 
+ (\Delta{\cal M}_{B}^2)_{A \alpha} ({\cal M}_{B}^2)_{\alpha \beta}^{-1} (\Delta{\cal M}_{B}^2)_{\beta B}
\right\}. 
\ea
As for the increment of the fermion mass matrix denoted by $\Delta {\cal M}_F,$ and $\Delta \overline{{\cal M}}_F$, 
we obtain
\ba
&&\Delta {\cal M}_F =
\left(
\begin{array}{cc}
0 & (\Delta {\cal M}_F)_{\lambda \psi} \\
(\Delta {\cal M}_F)_{\psi \lambda} & 0 \\
\end{array}
\right), \quad 
\Delta \overline{{\cal M}}_F = 
\left(
\begin{array}{cc}
0 & (\Delta \overline{{\cal M}}_F)_{\bar{\lambda} \bar{\psi}} \\
(\Delta \overline{{\cal M}}_F)_{\bar{\psi} \bar{\lambda}} & 0 \\
\end{array}
\right), \\
&&(\Delta {\cal M}_F)_{\psi \lambda} =  -\frac{\sqrt{2}}{2} i (\partial \hat{\fD})_*, \quad
(\Delta {\cal M}_F)_{\lambda \psi} = -\frac{\sqrt{2}}{2} i (\partial \hat{\fD})_*^T , \nonumber \\
&&(\Delta \overline{{\cal M}}_F)_{\bar{\psi} \bar{\lambda}} = \frac{\sqrt{2}}{2} i (\bar{\partial} \hat{\fD})_*, \quad
(\Delta \overline{{\cal M}}_F)_{\bar{\lambda} \bar{\psi}}  = \frac{\sqrt{2}}{2} i (\bar{\partial} \hat{\fD})_*^T. 
\ea

Let us now turn to the question of the mass sum rule. 
The increment of the bosonic part of the supertrace mass squared is
\ba
\Delta ({\rm Tr}M^2_{bosons}) &\equiv& {\rm tr} g_*^{-1} (\Delta {\cal M}_{B, red}^2)_{\bar{\phi} \phi}
+ {\rm tr} g_*^{-1} (\Delta {\cal M}_{B, red}^2)_{\phi \bar{\phi}}
+3 {\rm tr} ({\rm Im}{\cal F})_*^{-1}  (\Delta {\cal M}_B^2)_{VV} \nonumber \\
&=& 
-\frac{i}{4}{\rm tr} g_*^{-1}
\left\{
 (\partial {\cal F} \cdot D)_* ({\rm Im}{\cal F})_*^{-1} (\bar{\partial} \hat{\fD})_*^T  
+(\bar{\partial} \hat{\fD})_* ({\rm Im}{\cal F})_*^{-1}(\partial {\cal F} \cdot D)_*  
\right\} \nonumber \\
&& +\frac{i}{4}{\rm tr} g_*^{-1}
\left\{
(\partial \hat{\fD})_* ({\rm Im}{\cal F})_*^{-1} (\bar{\partial} \bar{{\cal F}} \cdot D)_* 
+(\bar{\partial} \bar{{\cal F}} \cdot D)_* ({\rm Im}{\cal F})_*^{-1} (\partial \hat{\fD})_*^T 
\right\} \nonumber \\
&& 
+ {\rm tr} g_*^{-1}
\left\{
(\partial \hat{\fD})_* ({\rm Im}{\cal F})_*^{-1} (\bar{\partial} \hat{\fD})_*^T
+(\bar{\partial} \hat{\fD})_* ({\rm Im}{\cal F})_*^{-1}(\partial \hat{\fD})_*^T  
\right\}. 
\ea
As for the trace of the fermion mass squared, the increment is
\ba
\Delta (2{\rm Tr} M^2_{fermions}) &=& 2{\rm tr} (\Delta{\cal M}_F) (G_F^{-1} \overline{\cal M}_F G_F^{-1}) 
+ 2 {\rm tr} {\cal M}_F (G_F^{-1} \Delta \overline{\cal M}_F G_F^{-1}) 
\nonumber \\
&&+ 2{\rm tr} (\Delta{\cal M}_F) (G_F^{-1} \Delta \overline{\cal M}_F G_F^{-1}) \nonumber \\
&=& +\frac{i}{2}{\rm tr} 
\left\{
(\partial \hat{\fD})_*^T g_*^{-1} (\bar{\partial} \bar{{\cal F}} \cdot D)_* ({\rm Im}{\cal F})_*^{-1} 
+ (\partial \hat{\fD})_* ({\rm Im}{\cal F})_*^{-1} (\bar{\partial} \bar{{\cal F}} \cdot D)_* g_*^{-1}
\right\} \nonumber \\
&& 
-\frac{i}{2}{\rm tr} 
\left\{
(\partial {\cal F} \cdot D)_* g_*^{-1} (\bar{\partial} \hat{\fD})_* ({\rm Im}{\cal F})_*^{-1} 
+ (\partial {\cal F} \cdot D)_* ({\rm Im}{\cal F})_*^{-1} (\bar{\partial} \hat{\fD})_*^T g_*^{-1} 
\right\} \nonumber \\
&& 
+ {\rm tr} \left\{
(\partial \hat{\fD})_*^T g_*^{-1} (\bar{\partial} \hat{\fD})_* ({\rm Im}{\cal F})_*^{-1} 
+ (\partial \hat{\fD})_* ({\rm Im}{\cal F})_*^{-1} (\bar{\partial} \hat{\fD})_*^T g_*^{-1}
\right\}. 
\ea
The increment of the supertrace is, therefore,
\ba
 && \Delta ({\rm Tr}M^2_{bosons} - 2{\rm Tr} M^2_{fermions})  \nonumber \\
&=& -{\rm tr} (g_*^{-1} D_* \cdot (\partial \bar{\partial} \hat{\fD})_*) \nonumber \\
 && +\frac{i}{4}{\rm tr} g_*^{-1}
\left\{
 (\partial {\cal F} \cdot D)_* ({\rm Im}{\cal F})_*^{-1} (\bar{\partial} \hat{\fD})_*^T  
+(\bar{\partial} \hat{\fD})_* ({\rm Im}{\cal F})_*^{-1}(\partial {\cal F} \cdot D)_*  
\right\} \nonumber \\
&& -\frac{i}{4}{\rm tr} g_*^{-1}
\left\{
(\partial \hat{\fD})_* ({\rm Im}{\cal F})_*^{-1} (\bar{\partial} \bar{{\cal F}} \cdot D)_* 
+(\bar{\partial} \bar{{\cal F}} \cdot D)_* ({\rm Im}{\cal F})_*^{-1} (\partial \hat{\fD})_*^T 
\right\}. 
\label{Deltasupertrace}
\ea
The quadratic piece in  the $D$ condensate being absent,
the right hand side is a generalization of the well-known expression $-{\rm tr} {\displaystyle \sum_a } D^a_* T^a$ 
 in the renormalizable supersymmetric gauge theories.
    The right hand side vanishes when the anomaly free property of the theory under concern
    is imposed. (See, for instance, \cite{Nilles}.)

 This completes the calculation which we have begun in section 3.
  To summarize,  the answer is given  by the two
  equations for the supertrace, eqs.(\ref{masssumrulesec4}) and (\ref{Deltasupertrace}).
  The right hand side of the mass sum rule vanishes by the special  K\"ahler geometry one adopts 
  and the anomaly free property of the theory under concern.

\section{Simple application of the mass sum rule}

In this section, we give a simple application of the mass sum rule derived above. 
For simplicity, we consider the situation of section 3, the mass sum rule for the sector consisting only of the
fields in the adjoint representation in the unbroken gauge group and the case in which the right hand side of
eq.(\ref{masssumrulesec4}) vanishes.
The mass sum rule for the vector multiplet and the adjoint chiral multiplet is given by 
\ba
(m_\phi^+)^2 + (m_\phi^-)^2  = 2((\Lambda^{(+)})^2 + (\Lambda^{(-)})^2)
\ea
where $m_{\phi}^{\pm}, m_\psi$ and $m_{\lambda}$  
 are adjoint scalar masses and 
$\Lambda^{\pm}$ are mass eigenvalues of mixed Majorana-Dirac fermions ( the mass eigenstates of the adjoint fermion
 mixed with the ordinary Majorana gaugino)
 obtained in \cite{IMaru3}
\ba
\Lambda^{(\pm)} = ({\rm tr}{\cal M})\lambda^{(\pm)},
\ea
where
\ba
{\cal M}_a &=&
\left(
\begin{array}{cc}
-\frac{i}{2} g^{aa} {\cal F}_{0aa} F^0, & -\frac{\sqrt{2}}{4} \sqrt{g^{aa} ({\rm Im}{\cal F})^{aa}}{\cal F}_{0aa} D^0 \\
-\frac{\sqrt{2}}{4} \sqrt{g^{aa} ({\rm Im}{\cal F})^{aa}}{\cal F}_{0aa} D^0, & g^{aa} \partial_a \partial_a W + g^{aa} g_{0a,a} \bar{F}^0 \\
\end{array}
\right) 
= \left(
\begin{array}{cc}
m_{\lambda\lambda}^a & m_{\lambda\psi}^a \\
m_{\psi \lambda}^a 
& m_{\psi\psi}^a \\
\end{array}
\right), \\
\lambda^{(\pm)} &=& \frac{1}{2} \left( 1 \pm \sqrt{(1+if)^2 + \left( 1+\frac{i}{2}f \right)^2 \Delta^2 } \right), \quad 
\Delta \equiv -  \frac{2m_{\lambda\psi}}{m_{\psi\psi}}, \quad f \equiv \frac{2im_{\lambda\lambda}}{{\rm tr}{\cal M}}.
\ea
From $(\Lambda^{(+)})^2 > 0$, we obtain an upper bound for the gaugino mass $\Lambda^{(-)}$
\ba
(\Lambda^{(-)})^2 < \frac{1}{2} [(m_\phi^+)^2 + (m_\phi^-)^2] = ({\cal M}_{red}^2)_{\phi\bar{\phi}}. 
\ea
In phenomenological applications, 
 it would be interesting to apply this relation to the gluino mass 
 since the lower bound for the gluino mass is severely constrained by the recent LHC data. 
Taking into account this lower bound, 
 we can predict an allowed range for the gluino mass as 
\ba
m_{\tilde{g}_{\rm lower~bound}} < m_{\tilde{g}} < ({\cal M}_{red})_{\phi\bar{\phi}}.  
\label{range}
\ea 
The scale $({\cal M}_{red})_{\phi\bar{\phi}}$ is naively given by a superpotential mass scale 
 $M_{sup} \sim (\partial_\phi \partial_\phi W)_* $, which must be much smaller than the cutoff (or the prepotential) scale 
 from the argument that lifetime of our metastable supersymmetry breaking vacuum should be longer
  than the age of the universe. This prediction eq.(\ref{range}) would be useful in  phenomenological study in LHC Run II.

\section{Soft SUSY breaking terms generated by the condensates}
In this section, we represent the mass and interaction terms generated by the condensates
in eq.(\ref{Lextended}) 
 as the supersymmetry breaking terms, using the spurion technique.

First, we notice that the background (spurion) fields in the present case are 
\ba
&&V_*^0 = \frac{1}{2} \theta^2 \bar{\theta}^2 D_*^0~{\rm or}~ {\cal W}_{\alpha*}^0 = \theta_\alpha D_*^0, \\
&&\Phi_*^0 = \varphi_*^0 + \theta^2 F_*^0 
\ea
and its conjugate. Exploiting these, the Lagrangian for these soft supersymmetry breaking terms is given by
\ba
{\cal L}_{soft} &=& \int d^4 \theta 
\left[
-\sum_{X=Q, U^*, D^*, L, E^*} \bar{X} e^{gV_*^0} X \right. \nonumber \\ 
&& \left. - \left\{
\frac{1}{M_{prep}^2}\bar{\Phi}_*^0 \Phi_*^0 + \frac{1}{M_{prep}^4} 
\left(
\Phi_*^0 \overline{{\cal W}}_*^0 \overline{{\cal W}}_*^0 + \bar{\Phi}_*^0 {\cal W}_*^0 {\cal W}_*^0
\right) 
+ \frac{1}{M_{prep}^6} \overline{{\cal W}}_*^0 \overline{{\cal W}}_*^0 {\cal W}_*^0 {\cal W}_*^0
\right\}
\sum_{X=Q, U^*, D^*, L, E^*} \bar{X}X
\right] \nonumber \\
&&-
\left[
\int d^2 \theta 
\left\{
\frac{1}{M_{prep}} {\rm tr} ({\cal W}_*^0 \Phi^a {\cal W}^a) 
+ \left( \frac{1}{M_{prep}} \Phi_*^0 + \frac{1}{M_{prep}^3} {\cal W}^0_* {\cal W}^0_* \right) y_u Q U^* H_u 
\right. \right. \nonumber \\
&& \left. \left. 
+ \left( \frac{1}{M_{prep}} \Phi_*^0 + \frac{1}{M_{prep}^3} {\cal W}^0_* {\cal W}^0_* \right) y_d Q D^* H_d 
+ \left( \frac{1}{M_{prep}} \Phi_*^0 + \frac{1}{M_{prep}^3} {\cal W}^0_* {\cal W}^0_* \right) y_e L E^* H_d \right. \right. \nonumber \\
&& \left. \left. 
+\left(
\Phi_*^0 + \frac{1}{M_{prep}^2} {\cal W}_*^0 {\cal W}_*^0
\right) H_u H_d
-\left( \frac{1}{M_{prep}} \Phi_*^0 + \frac{1}{M_{prep}^3} {\cal W}_*^0 {\cal W}_*^0 \right) 
{\rm tr} ({\cal W}^a {\cal W}^a)
\right\} +{\rm h.c.}
\right]
\label{softterms}
\ea
where $X=Q, U^*, D^*, L,$ and  $E^*$ denote the SM chiral multiplets, 
 ${\cal W}^a$ the SM gauge field strength, 
 $\Phi^a$ adjoint chiral multiplets with the SM charges. We have simply omitted
${\cal O}(1)$ coefficients of the operators. 

The terms in the first and the second lines in  eq.(\ref{softterms}) generate the scalar masses after supersymmetry breaking.   
The first term of the third line in eq.(\ref{softterms}) is a term to generate Dirac gaugino mass. 
The remaining terms in the third and the fourth lines are A-terms, 
 and the first terms in the last line of eq.(\ref{softterms}) represent $B\mu$ term of soft supersymmetry breaking terms.  
The last terms in the last line of eq.(\ref{softterms})  generates Majorana gaugino masses. The 
$\mu$-term, which is supersymmetric, can be obtained as well by the VEV of $\Phi^0_*$ from the first term 
in the last line of eq.(\ref{softterms}). 

This Lagrangian (\ref{softterms}) written in terms of spurion superfields of soft SUSY breaking terms is expanded in components, 
\ba
{\cal L}_{soft} &=& -\sum_{X=Q, U^*, D^*, L, E^*} m_{\tilde{X}}^2 \bar{\tilde{X}} \tilde{X} \nonumber \\
&& + \left[
- m_D \lambda^a \psi^a 
- A_u \tilde{Q} \tilde{U^*} H_u 
- A_d \tilde{Q} \tilde{D^*} H_d 
- A_e \tilde{L} \tilde{E^*} H_d 
+ B\mu H_u H_d -m_M \lambda^a \lambda^a + {\rm c.c.}
\right] \nonumber \\
\ea
where the fields with tilde represent the scalar component of the corresponding chiral superfield, 
and the parameters for each operators are provided in terms of $D_*^0$ and $F_*^0$ as 
\ba
m_{\tilde{X}}^2 &=& \frac{g}{2} D_*^0 + \frac{1}{M_{prep}^2} \bar{F}_*^0 F_*^0 
+ \frac{1}{M_{prep}^4} \left(F_*^0 (D_*^0)^2 + \bar{F}_*^0 (D_*^0)^2 \right) + \frac{1}{M_{prep}^6} (D_*^0)^4 \nonumber \\
&\sim& \frac{g}{2} M_{sup} M_{prep} + \frac{1}{M_{prep}^2} \bar{F}_*^0 F_*^0 
+ \frac{M_{sup}^2}{M_{prep}^2} \left(F_*^0 + \bar{F}_*^0 \right) + \frac{M_{sup}^4}{M_{prep}^2}, \\
m_D &=& \frac{1}{M_{prep}} D_*^0 \sim M_{sup}, \\
A_{u,d,e} &=& \frac{y_{u,d,e}}{M_{prep}} F_*^0 + \frac{y_{u,d,e}}{M_{prep}^3} (D_*^0)^2 
\sim \frac{y_{u,d,e}}{M_{prep}} F_*^0 + \frac{y_{u,d,e}}{M_{prep}} M_{sup}^2, \\
B\mu &=& F_*^0 + \frac{1}{M_{prep}^2} (D_*^0)^2 \sim F_*^0 + M_{sup}^2, \\
m_M &=& \frac{1}{M_{prep}} F_*^0 + \frac{1}{M_{prep}^3} (D_*^0)^2 \sim \frac{1}{M_{prep}} F_*^0 + \frac{M_{sup}^2}{M_{prep}} 
\ea
where $D^0_* \sim M_{sup} M_{prep}$ is put in the final expressions.  


\section*{Acknowledgments}
The work of H.I. is supported in part by the Grant-in-Aid 
 for Scientific Research from the Ministry of Education, 
 Science and Culture, Japan No. 23540316.
The work of N.M. is supported in part by the Grant-in-Aid 
 for Scientific Research from the Ministry of Education, 
 Science and Culture, Japan No. 24540283. 

\newpage

\appendix

\section*{Appendix}

In the text, we have introduced the notation in eqs.(\ref{bosonmass})-(\ref{fermionmassbar}) 
 such that our computation as well as the mass sum rule is insensitive to 
 the structure of vacua explored. 
In a specific vacuum one works with, these expressions get further simplified 
 but become noncovariant. 
For instance, in the unbroken vacuum of the $U(N)$ gauge group, 
 the nonvanishing entries of ${\cal F}_{abc*}, D^a_*, F^a_{*}$ are 
 \ba
 {\cal F}_{0aa*}={\cal F}_{000*}, \quad 
D^a_*=\delta_0^a D^0_*, \quad 
F_*^a=\delta_0^a F_*^0, \quad 
 g_{a\bar{b}}=\delta_{\bar{b}\bar{a}} g_{a\bar{a}}= \delta_{b\bar{a}}g_{00}. 
\ea
Consequently, 
\ba
\left( F\cdot \bar{\partial} \partial g \cdot \bar{F} \right)_{*\bar{a}b} = \bar{F}_*^0 g_{00\bar{a}b*} F_*^0, \qquad
\left( (\partial {\cal F} \cdot ) D \right)_{*ab} = \delta_{ab} {\cal F}_{000*} D_*^0, 
{\rm etc.}
\ea
For more complex cases such as $U(N)$ is broken to product groups, 
 see, for instance, \cite{FIS3}. 

\newpage


\end{document}